\documentclass[prb,aps,twocolumn,superscriptaddress,10pt]{revtex4-1}
\usepackage{graphicx}
\usepackage{amsmath,amssymb,amsfonts}

\usepackage{color}
\definecolor{darkblue}{rgb}{0.,0.,0.4}
\definecolor{darkred}{rgb}{0.5,0.,0.}
\usepackage[pdftex,colorlinks=true,linkcolor=darkblue,citecolor=darkred,urlcolor=blue]{hyperref}

\newcommand{\bra}[1]{\left\langle #1 \right|}
\newcommand{\ket}[1]{\left| #1 \right\rangle}
\newcommand{\braket}[2]{\left\langle#1 \middle|  #2\right\rangle}
\newcommand{\rd}[1]{\mathop{\mathrm{d}#1}}
\newcommand{\abs}[1]{\left| #1 \right|}

\begin{document}
\title{Localization from superselection rules in translation invariant systems}

\date{\today}

\author{Isaac H. Kim}
\affiliation{Perimeter Institute for Theoretical Physics, Waterloo, Ontario, Canada}
\affiliation{Institute for Quantum Computing, University of Waterloo, Waterloo ON N2L 3G1, Canada}
\author{Jeongwan Haah}
\affiliation{Department of Physics, Massachusetts Institute of Technology, Cambridge, Massachusetts, USA}

\begin{abstract}
We study a translation invariant spin model in a three-dimensional regular lattice, called the cubic code model,
in the presence of arbitrary extensive perturbations.
Below a critical perturbation strength, we show that most states with finite energy are localized;
the overwhelming majority of such states have energy concentrated around a finite number of defects,
and remain so for a time that is near-exponential in the distance between the defects.
This phenomenon is due to an emergent superselection rule and does not require any disorder.
An extensive number of local integrals of motion for these finite energy sectors are identified as well.
Our analysis extends more generally to systems with immobile topological excitations.
\end{abstract}

\maketitle

\section{Introduction}
Recently there has been a significant amount of interest on the mechanism behind how isolated quantum systems thermalize.
While the eigenstate thermalization hypothesis
asserts that quantum many-body systems thermalize at the level of the eigenstates,\cite{Deutsch1991,Srednicki1994}
there are well-known counterexamples to this proposal as well.
In particular, a seminal work of Anderson shows that disorder
can inhibit thermalization for noninteracting systems,
as particles are localized indefinitely at some fixed location.\cite{AndersonLocalization1958}
Recent studies indicate that the effect of localization
can persist in the presence of interaction, even at finite energy densities.%
\cite{Basko2006,Pal2010,LocalConservationLawMBL2013,Huse2013}
These results suggest that disorder can cause the system to ``freeze'' in time, and hinders equilibration.
There are also some translation invariant systems
which were proposed to be localized under a generic disordered initial conditions.%
\cite{QuantumDisentangledLiquids2013,Schiulaz2013,Roeck2013,Hickey2014,Schiulaz2014,Yao2014}
However, currently there does not seem to be a decisive consensus
as to whether such effects persist in the thermodynamic limit;
see Ref.~\onlinecite{Papic2015}.

In this paper, we point out that interacting quantum many-body systems
can be localized by a completely different mechanism.
This is due to an emergent superselection rule, and does not rely on disorder.
We consider an exactly solvable and translation-invariant lattice model of spins, called the cubic code model\cite{Haah2011Local},
and find a set of states with manifestly localized spatial energy profile.
These states remain invariant under Hamiltonian evolution,
and constitute the overwhelming majority of the states at finite energy.
Our main result here is that all these properties remain stable against {\em arbitrary perturbations},
so long as the perturbation strength is smaller than some finite critical value.
Hence we conclude that this phenomenon is a robust property of the phase.

To be more precise,
we consider a weak arbitrary perturbation, denoted as $Y$, to the cubic code model $H_0$ that shall be defined later,
and construct a set of states $\ket{\psi_Y}$
such that their energy is concentrated around a finite number of points
with respect to the perturbed Hamiltonian $H_0 + Y$.
These states are quasi-eigenstates in the sense that
\begin{equation}
 \abs {\bra{\psi_Y} e^{-it(H_0+Y)} \ket{\psi_Y} } \ge 1 - t L^\alpha \exp( - c L^\eta )
 \label{eq:rough-result}
\end{equation}
where $\alpha, c, \eta > 0$ are constants and $L$ is the smallest distance between the defects.
Furthermore, the fraction of these states within the low-energy subspace
is shown to approach $1$ in the thermodynamic limit.

While we only consider a system with finite number of defects,
we emphasize that our system is fundamentally different from noninteracting fermionic systems.
The transport of the low-energy excitations are hindered by a novel form of emergent superselection rule,
which does not arise in free systems.
Interestingly, localization originates from strong interaction within our setup,
contrary to what is observed in the context of disordered systems.

It should be noted that the localized states satisfying Eq.~\eqref{eq:rough-result}
are unlikely to exist in a system with nontrivial dispersion relation.
Consider a one-dimensional fermionic chain with a Hamiltonian
$H=\int \rd k \epsilon(k) c^\dagger_k c_k$ as an example.
Here, $c_k$ is the annihilation operator of momentum $k$.
If the dispersion relation is $\epsilon(k) = -\lambda \cos k$,
then a state representing a single localized excitation,
such as $\ket \psi = \int \frac{\rd k}{\sqrt{2\pi}} c^\dagger_k \ket \Omega$,
will spread out quickly in time;
\begin{align}
 \bra \psi e^{-itH} \ket \psi &=  J_0(\lambda t) = 1 - \frac{\lambda^2 t^2}{4} + \cdots, \nonumber
\end{align}
where $J_0(t)$ is the Bessel function.

In the rest of the paper, we shall introduce our model and explain how Eq.~\eqref{eq:rough-result} is derived.
We first begin by explaining the central concept, the \emph{locally gapped state.}
Roughly speaking, the notion of locally gapped state formalizes the intuition
that two eigenstates do not mix with each other with respect to a perturbation
if they are either (i) separated from each other with a large energy gap
or (ii) the corresponding off-diagonal matrix element of the perturbation is sufficiently small.
We then proceed by identifying all the low-energy states of our model which are locally gapped.
A simple estimate shows that locally gapped states form the overwhelming majority of the states within the low-energy subspace.
Lastly, we rigorously prove that the aforementioned properties remain stable in the presence of weak perturbations,
and find an extensive set of approximate local integrals of motion.
Throughout the paper, we shall adopt the following technical terminology:
A  function $f(x)$ {\em rapidly decays}
if it scales like $x^\alpha \exp( -c x^\beta )$ for large $x$,
where $c, \beta> 0 $ and $\alpha$ are constants.

\section{Locally gapped states}

With respect to a Hamiltonian $H$,
a state $\ket \psi$ is said to be {\em locally gapped}
with a diameter $d$ and an energy $\Delta$,
if for any orthogonal state $\ket \phi$, one of the following holds:
\begin{itemize}
 \item $\abs{ \bra \phi H \ket \phi - \bra \psi H \ket \psi } \ge \Delta >0$
 \item $\bra \phi O \ket \psi = 0$ whenever $O$ is supported on a ball of diameter $d$.
\end{itemize}
\noindent
This means that $\ket \psi$ is separated by an energy gap of $\Delta$
from all the other states that are reachable by local operators from $\ket \psi$.
Unless specified otherwise, we will be interested in locally gapped states with an extensive diameter and finite energy gap.
Let us mention some examples.
A non-degenerate ground state of a Hamiltonian with an energy gap is locally gapped.
A less trivial example is the ground state sector of topologically ordered systems.
The ground states can be only connected to each other by an operator whose support is extended across the system.

For clarity, let us also discuss states which are {\em not} locally gapped.
Consider a Hamiltonian $H = \sum_i \sigma^z_i$ on a chain where $\sigma^z$ is the Pauli-$z$ matrix.
It is gapped with a unique ground state in which all the spins are pointing ``down.''
Its first excited state is a spin configuration with a single spin, say the $j$th one, pointing up.
This excited state is not locally gapped because there is another excited state for which only the $(j+1)$th spin is pointing up.
These two are related to each other by a local operator, namely $\sigma^x_j \sigma^x_{j+1}$.
These states are not locally gapped with diameter $d >1$.

The above conclusion exemplifies a general observation:
If the hopping of a quasi-particle is realized by a local operator, states with localized excitations are not locally gapped.
This intuition is true for many translation-invariant systems,
and it is thus tempting to conjecture that the only way to have a locally gapped  excited state
is through some strong disorder.
However, we shall show that there exists a translation-invariant system with many locally gapped excited states.

One might want to consider an approximate and more general version of locally gapped states
by allowing the transition amplitude $\bra \phi O \ket \psi$ to be approximately equal to zero for local operators $O$.
We shall encounter this generalization later, but for now it will only make the notation more cumbersome.

\section{Locally gapped states in the pure cubic code Hamiltonian}

\begin{figure}[t]
\includegraphics[trim = 0ex 1ex 0ex 1ex, clip = true, width=.45\textwidth]{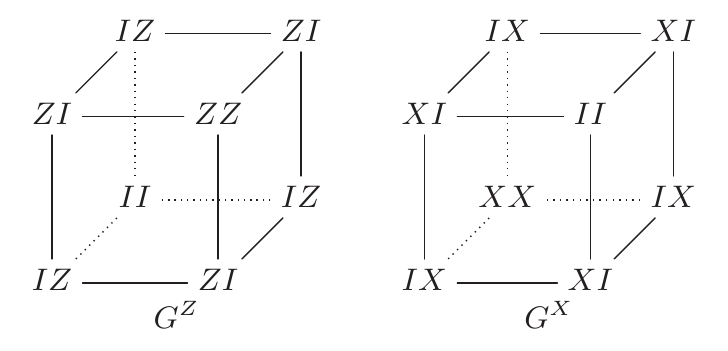}
\caption{Cubic code model. $X$ and $Z$ represent Pauli matrices $\sigma^x$ and $\sigma^z$, respectively.
A term in the Hamiltonian is the product of 8 Pauli matrices arranged as in the diagram.}
\label{fig:cubicCode}
\end{figure}

The cubic code model is defined on a simple cubic lattice with two qubits per site,
where the Hamiltonian is a translation-invariant sum of the two types of interaction terms~\cite{Haah2011Local}.
\begin{align}
 H_0 &= - J\sum_i G^Z_i + G^X_i \quad \quad (J >0) \label{eq:cubic_code}\\
 G^Z_i &= \sigma^z_{i,1} \sigma^z_{i,2} \sigma^z_{i-\hat x,1} \sigma^z_{i-\hat y,1} \sigma^z_{i-\hat z,1}
         \sigma^z_{i-\hat y-\hat z,2} \sigma^z_{i-\hat z-\hat x,2} \sigma^z_{i-\hat x-\hat y,2} \nonumber \\
G^X_i &= \sigma^x_{i,1} \sigma^x_{i,2} \sigma^x_{i+\hat x,2} \sigma^x_{i+\hat y,2} \sigma^x_{i+\hat z,2}
         \sigma^x_{i+\hat y+\hat z,1} \sigma^x_{i+\hat z+\hat x,1} \sigma^x_{i+\hat x+\hat y,1} \nonumber
\end{align}
See Fig.~\ref{fig:cubicCode} for the arrangement of the Pauli marices in the terms.
Every term in the Hamiltonian commutes with each other, and as such, the energy spectrum is discrete.
For the ground state, both $G^X_i$ and $G^Z_i$ take eigenvalues of $+1$.
The ground state subspace is degenerate and is topologically ordered in the sense that any local operator
has the same expectation value for all the ground states.
According to our definition in the previous Section,
the ground state is locally gapped by an energy $\Delta = 2J$ and diameter $L-1$,
where $L$ is the linear system size.
The excited states can be described by {\em defects}, which are the violated local terms, e.g., $G^X_i = -1$ or $G^Z_i = -1$.
We refer to each types of the violated terms as $X$-type and $Z$-type.
Since $X$-type term and $Z$-type term are lattice inversions of each other, it suffices to analyze just one of the two types.
Not every configuration of defects is physically allowed. Indeed, a configuration of defects can be realized as an excited state
if and only if it can be created by applying a tensor product $P$ of Pauli matrices (Pauli operators) to the ground state.
We call such defect configurations to be {\em valid.}

The most general energy eigenstate is a superposition of valid configurations with the same number of defects.
Note that the configuration of defects does not uniquely determine the state.
For example, the ground state has no defect, but is degenerate.
Given the defect configuration, this is in fact the only residual degeneracy.
We shall call this as the {\em topological degeneracy}.

Below, we will make use of the following fact\cite{Haah2012PauliModule}.
Under periodic boundary conditions there is an infinite family of system sizes
such that any defect configuration with an even number of $X$-type defects
and an even number of $Z$-type defects is valid, and the topological degeneracy is $4$.
This is the family that we shall study in this paper.
Concretely, it includes the lattices of linear size $L = 2^n+1$ for any integer $n \ge 1$.
The situation is very similar to the toric code Hamiltonian;
one cannot have an isolated anyon in a closed finite system,
but any even number of anyons is allowed.

\subsection{No-strings rule}

The most important property, called {\em no-strings rule}\cite{BravyiHaah2011Energy},
is a formalization of the fact that isolated defects are immobile.
More precisely, suppose $\ket \psi$ is a state with a defect at site $i$ and no other defects within distance $d$ from $i$.
If $\ket{\psi'}$ is another state with a defect at $i' \neq i$ and no other defect within distance $d$ from $i$,
then the no-strings rule asserts that
$\bra{\psi'} O \ket \psi = 0$ for any operator $O$ supported on a ball of diameter $d$.
Note that the no-strings rule is a nontrivial condition, in that $i'$ can be even in the vicinity of $i$.

\begin{figure}[b]
\includegraphics[trim=10ex 55ex 10ex 55ex, clip=true, width=.45\textwidth]{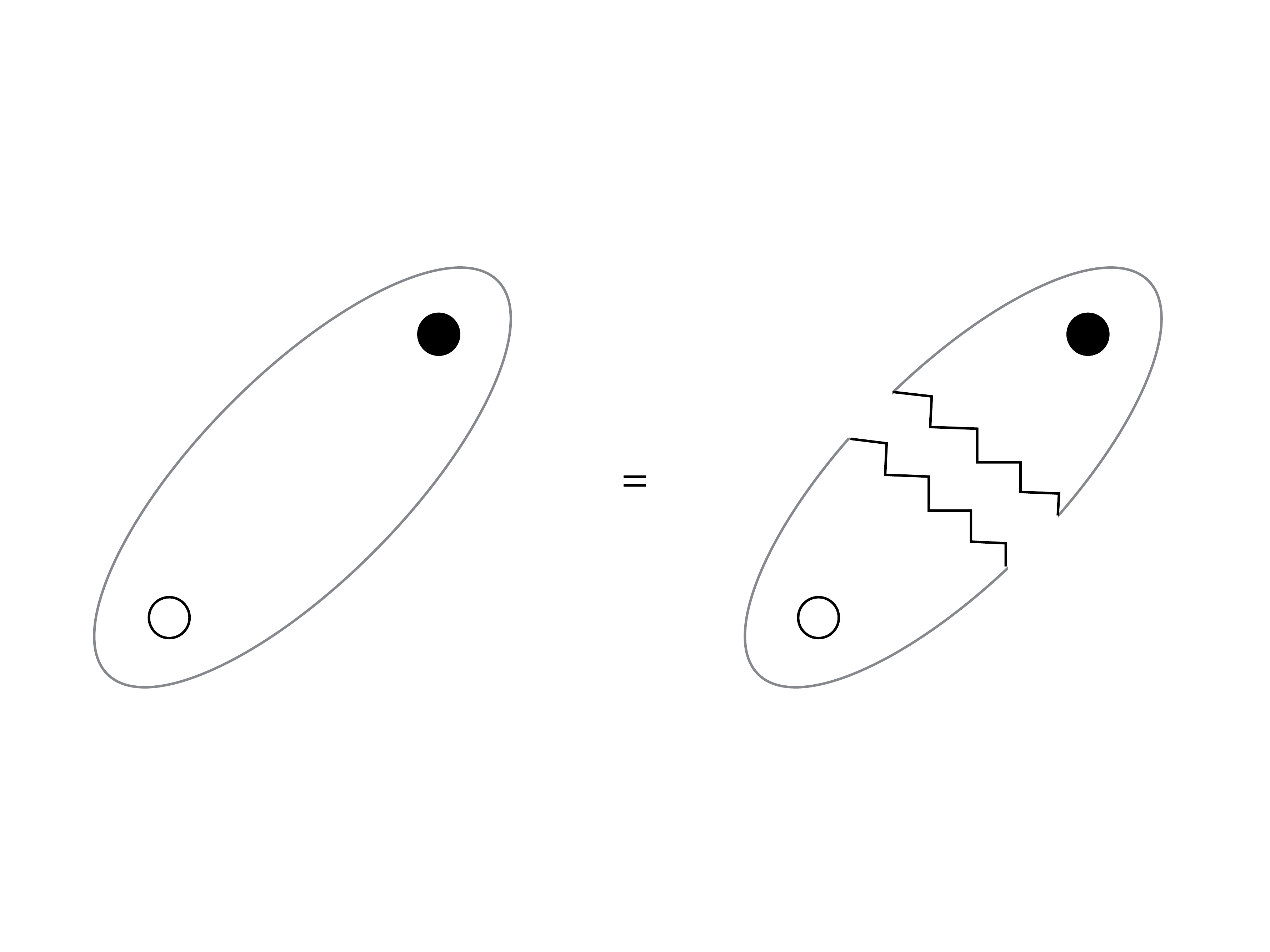}
\caption{No-strings rule.
If some operator supported on a bounded region of the lattice
can transport a defect, then there exists another operator
that has exactly the same effect on the state and is supported on the vicinity of the initial and final defects.
This is impossible since a single defect cannot be locally created or annihilated.
Therefore, no such operator exists.
See Ref.~\onlinecite{Haah2011Local} for more detail.}
 \label{fig:no-strings-rule}
\end{figure}

The cubic code satisfies the no-strings rule, as we explain below.
Note that any operator can be written as a linear combination of tensor products of Pauli matrices (Pauli operators).
This is possible simply because Pauli matrices generate the full operator algebra.
This observation eases the problem; in order to argue that no operator can ``transport'' a defect,
it suffices to show that no Pauli operators can transport the defect.
If a Pauli operator $O$ supported on a ball of diameter $d$
is capable of ``transporting'' a defect within that ball,
some product of $O$ and its translations can transport the defect arbitrarily far.
The calculation of Ref.~\onlinecite{Haah2011Local} implies that, if a defect  can be transported arbitrarily far,
it can be equally created from the ground state by acting some Pauli operator in its vicinity; see Fig.~\ref{fig:no-strings-rule}.
This is impossible since a local Pauli operator can only create an even number of defects.
We arrive at a contradiction, so we conclude that any Pauli operator, and hence any operator,
is incapable of transporting defects.

The no-strings rule implies the existence of low-energy excited states which are locally gapped.
To see this, consider an excited state $\ket \psi$ describing a configuration of defects
that are separated from one another by a distance $d$.
Let $e$ be the the total number of defects, so that the energy of the state is $2Je$.
If another state of the same energy, denoted as $\ket \phi$, has a configuration of defects different from that of $\ket \psi$,
there must be a defect at $i$ that is present in $\ket \phi$ but not in $\ket \psi$.
By the construction of $\ket \psi$, the no-strings rule implies that the matrix element $\bra{\phi} O \ket \psi$ is zero
for any operator $O$ whose support is contained in a ball of diameter $d$.
More generally, any state orthogonal to $\ket \psi$ with the same energy can be labeled by two components.
One represents different defect configurations with the same energy,
and the other represents topological degeneracy.
The transition from $\ket \psi$ to any orthogonal and topologically degenerate state
requires an operator whose support is comparable to the system size.
By the no-strings rule, the transition from $\ket \psi$ to any other state with a different defect configuration
requires an operator whose support is at least $d$.
Other states have energy that is different from $2Je$ by at least $2J$.
Therefore, the no-strings rule implies that a configuration of separated defects by distance $d$ is locally gapped
with diameter $d$ and energy $2J$.

Remark that the existence of the locally gapped excited states is not due to any disorder;
it follows from the no-strings rule, which is a nontrivial property of the cubic code model.
A correct interpretation should be that defects at different locations represent distinct superselection sectors.
A similar yet weaker version of this phenomenon is observed in Wen's plaquette model~\cite{Wen2003Plaquette}.
There, a defect can be transported to next-nearest neighboring site by a local operator, but not to the nearest neighboring site.
In other words, the superselection sector is changed under a unit translation,
although it is not changed under two units of translations.
In the cubic code model, there is no such finite periodicity,
and consequently, there are infinitely many superselection sectors.

\subsection{Typicality of locally gapped states}

Now we show that almost all defect configurations are locally gapped.
Without loss of generality, let us  assume that $m$ defects are all $X$-type,
as the analysis for the $Z$-type defects are exactly the same.
Any configuration is valid as long as $m$ is even,
and there is no restriction on the positions of the defects.
The total number of states with a fixed energy will be given by
the number of distinct configurations, multiplied by the topological degeneracy $4$.
This is a purely combinatorial problem
in which one chooses $m$ defect locations from a lattice of $V$ sites
such that every defect is separated from the others by distance $d$.
Let us imagine that we are placing one defect at a time, eventually $m$ defects in total.
The first defect can be anywhere, as there are $V$ possibilities.
The second one can be anywhere except near the first one;
there are $V-v$ possibilities, where $v$ is the volume of a ball of radius $d$.
The third one can be placed anywhere among at least $V-2v$ positions.
Continuing in a similar way, we see that the number of these sparse defect configurations is at least
$ V(V-v)(V-2v) \cdots (V- (m-1)v) / m!$.
The total number of all configurations of defects is $\binom{V}{m}$.
Thus the fraction of the locally gapped states with diameter $d$ is at least
\begin{equation}
 \frac{ V(V-v) \cdots (V- (m-1)v) }{V(V-1) \cdots (V-m+1)} \ge \left( 1 - \frac{mv}{V} \right)^m .
 \label{eq:fractionlocally gapped}
\end{equation}
The fraction approaches $1$ algebraically in the system size,
provided that $d \sim L^{1-\epsilon}$ where $0 < \epsilon < 1$.
We conclude that the majority of the defect configurations is locally gapped.

There are excited states which are not locally gapped,
but they form a vanishing fraction of the low-energy subspace due to the previous analysis.
These are the excited states with locally created clusters of defects.
A local operator creates at least 4 defects from the ground state,
and this cluster of 4 defects can hop to a nearby position by local operators.
Thus, excited states with a topologically neutral (i.e., locally created) cluster of defects are not locally gapped,
and may hybridize to become a momentum eigenstate upon perturbations.

\section{locally gapped states under perturbations}

Locally gapped states are resilient to perturbations.
Before we see why, let us recall a basic fact from perturbation theory to develop some intuitions.
Consider a nondegenerate eigenstate $\ket{\psi}$ of some Hamiltonian $H$ with an energy $E$.
Under an infinitesimal perturbation $Y$, the eigenstate is perturbed as follows:
\begin{equation}
\delta \ket{\psi} = \sum_j \frac{\bra{\psi_j}Y \ket{\psi}}{E-E_j} \ket{\psi_j}, \nonumber
\end{equation}
where $\ket{\psi_j}$ are the energy eigenstates of $H$ with energy $E_j$.
The correction term does not blow up
if the state is either separated from the rest of the spectrum by a constant energy gap
or the off-diagonal matrix elements are sufficiently small.
These are exactly the two criteria that appear in the definition of locally gapped states.
If the first criterion is satisfied, the state evolves under an adiabatic evolution,
and we expect the wave function to change smoothly.
If the second criterion is satisfied, the perturbation acts almost diagonally with respect to the original energy eigenstates,
so we expect the effect of the perturbation to the eigenstates to be negligible.
These observations suggest that if the energy profile of $\ket \psi$ is localized
in the sense that there are a few positions around which energy density is nonzero,
then the perturbed state also has a similar energy profile.
This rough intuition will be made more precise in this Section.

\subsection{Band-preserving perturbations}
Consider a one-parameter family of Hamiltonians of the following form
\[
H_s=H_0+sY, \quad s \in [0,1],
\]
where $H_0$ is the original cubic code Hamiltonian (Eq.~\ref{eq:cubic_code}) and $Y$ is some perturbation.
We assume that the perturbation consists of local terms with bounded strength,
where strength of the perturbation refers to the supremum of the norms of the local terms.

Let us invoke a powerful theorem which is due to Bravyi, Hastings, and Michalakis~\cite{BravyiHastingsMichalakis2010Stability,BravyiHastings2011short}.
Setting the energies of $H_0$ to be $E_n$ ($n=0,1,2,\ldots$) with the ground state energy being $E_0 = 0$,
the theorem says that the spectrum of $H_s$ is contained in $[E_n(1-cs), E_n(1+cs)]$
up to a small correction that vanishes in the thermodynamic limit.
Here, the number $c>0$ is a constant that is independent of the system size,
and the strength of the perturbation is assumed to be smaller than some constant.
At finite energies, this implies that there are energy bands that are separated from each other by some constant, say $\Delta$.
This spectral gap condition allows us to unambiguously define projectors
onto the linear span of states within one of the separated bands.
We shall refer to such projectors as the projector onto the \emph{band subspace.}

\subsection{Evolution of the band subspace}
The existence of the separated energy bands at finite energy implies that
we can use a powerful machinery of quasi-adiabatic continuation~\cite{HastingsWen2005Quasi-adiabatic,Osborne2007}.
This allows us to systematically track the effect of the perturbation on these subspaces.
It says that, whenever a band subspace around energy $E_n$ is separated from the rest of the spectrum by a constant gap,
the projector $P_s^{(n)}$ onto the band subspace obeys the following differential equation:
\begin{equation}
 \partial_s P_s^{(n)} = i[K(s), P_s^{(n)}] ,
 \label{eq:qadiabatic-continuation}
\end{equation}
for some hermitian operator $K(s)$.
Formally, the solution to this equation is the following:
\begin{align}
 P_s^{(n)} &= U_s P_0^{(n)} U_s^\dagger \label{eq:UPU} \\
 U_s &= \mathcal T \exp \left( i \int_0^s K(u) \rd u \right)
\end{align}
where $\mathcal T$ is the ``time''-ordering operation.
The readers may wonder about the definition of $K(s)$ at this point.
While its rather esoteric definition shall be given in the Appendix,
its essential property can be summarized succinctly as follows.

The important point is that the unitary $U_s$ preserves locality.
Namely, for any operator $O$ supported on a region $M$, we have the following decomposition:
\begin{equation}
 U_s^\dagger O U_s = \sum_r O'_r, \quad \| O'_r \| \le  f(r),
 \label{eq:localityPreserving}
\end{equation}
where $O'_r$ is some operator supported on a distance-$r$ neighborhood of $M$ and $f(r)$ is some function which decays rapidly in $r.$

Note that $U_s$ governs the evolution of any energy band
which is separated from the rest of the spectrum by a constant gap.
As we have already discussed,
the energy bands of the perturbed cubic code at finite energy satisfies this condition,
provided that the strength of the perturbation is sufficiently small.

\subsection{Approximately locally gapped states}
By making use of these results, we can explicitly construct a large set of  (approximately) locally gapped states.
Such states are defined as follows:
\begin{equation}
\ket{\psi_s} = U_s \ket{\psi_0},
\end{equation}
where $\ket{\psi_0}$ is any state describing a sparse defect configuration of $H_0$,
the \emph{unperturbed} cubic code Hamiltonian.
Let us assume that
(i) the perturbation strength is sufficiently small so that $\ket{\psi_0}$
belongs to some separated band, say $P^{(m)}_s$, and
(ii) $\ket{\psi_0}$ is locally gapped with a diameter $d$.

These states are approximately locally gapped in the following sense.
For the states that are not in the band, there is an energy gap $\Delta$.
For the states in the band, we need to consider the transition amplitude.
Suppose there is an operator $O$ supported on a ball of diameter $d/2$ such that
$\bra{\phi_s} O \ket{\psi_s} = \delta$ where $\braket{ \phi_s}{\psi_s} = 0$.
This is to say that $\bra{\phi_0} U_s^\dagger O U_s \ket{\psi_0} = \delta$
where $\ket{\phi_0} = U_s^\dagger \ket{\phi_s}$ is a state that belongs to $P^{(m)}_0$.
Using the decomposition in Eq.~\eqref{eq:localityPreserving},
the operator $U_s^\dagger O U_s$ is quasi-local and hence can be approximated by an operator $O'$ supported on
the ball of diameter $d$ such that $\| U_s^\dagger O U_s - O' \|$ decays rapidly in $d$.
Since $\ket{\psi_0}$ is locally gapped with radius $d$,
$\bra{\phi_0} O' \ket{\psi_0} = 0$.
It follows that $\delta$ decays rapidly with $d$.
This establishes that $\ket{\psi_s}$ is approximately locally gapped with a diameter of $d/2$,
up to an ``error'' that decays rapidly with $d$.

\section{Properties of locally gapped states}

In the previous section, we have identified the locally gapped states of the perturbed Hamiltonian($H_s$).
In this section, we study their static and dynamical properties; we show that
they have spatially localized energy profile,
obey an area law of entanglement entropy,
behave almost like energy eigenstates of $H_s$,
and admit an extensive number of approximate local integrals of motion.
Since the ratio of the number of localized states relative to the dimension of the low-energy subspace is very close to 1
from our counting argument,
we conclude that a typical state from the low-energy subspace of arbitrarily perturbed Hamiltonian
consists of ``frozen'' defects.

\subsection{Static properties}

Here, we show that the spatial energy profile of $\ket{\psi_s}$ is localized.
We actually make a stronger claim: that the local reduced density matrix of $\ket{\psi_s}$ at a region $M$ is
close to that of the ground state of $H_s$ if $M$ is far from the locations of the defects for $\ket{\psi_0}$.

In order to see this, recall that the local reduced density matrix
is completely determined by the expectation values of local observables.
\begin{equation}
\bra{\psi_s} O\ket{\psi_s} = \bra{\psi_0} U_s^{\dagger}OU_s\ket{\psi_0}. \nonumber
\end{equation}
Since $U_s$ preserves locality, the ``dressed'' operator $U_s^{\dagger}OU_s$
can be well-approximated by some local operator, say $O'$.
If this local approximation acts away from the defects,
its expectation value with respect to $\ket{\psi_0}$
reproduces the ground state expectation value of the unperturbed system.
If the ground states of the unperturbed and perturbed Hamiltonians
are $\ket{\Omega_0}$ and $\ket{\Omega_s}$, respectively,
the preceding discussion implies that
$\bra{\psi_s} O \ket{\psi_s} \simeq \bra{\Omega_0} O' \ket{\Omega_0}
= \bra{\Omega_s} U_s O' U_s^\dagger \ket{\Omega_s} \simeq \bra{\Omega_s} O \ket{\Omega_s}$.
Therefore, any observable acting far from the defects cannot distinguish $\ket{\psi_s}$ from $\ket{\Omega_s}$.
The error term decays rapidly in the distance from $O$ to the defect's location.
(Recall that a rapid decaying function scales like $x^\alpha e^{-c x^\beta}$ for some constant $c, \beta >0$ and $\alpha$ for large $x$.)
In comparison, it is worth noting that the difference between the local reduced density matrices
of a momentum eigenstate and that of the ground state can only be algebraically small
($x^{-\gamma}$ for some $\gamma > 0$) in the system size.

We note in passing that the state $\ket{\psi_s}$ obeys an area law of entanglement entropy.
Namely, the entanglement entropy $S(M,\ket{\psi_s})$ for a ball-like region $M$
scales like the area of the boundary of $M$.
This is not too surprising in light of a recent result~\cite{VanMarienVerstraete2013};
in fact, it has nothing to do with our construction of states with a localized energy profile.
This is rather a consequence of the locality-preserving property of $U_s$
and the smallness of the low energy band subspace.
Consider first the unperturbed Hamiltonian $H_0$.
Every state $\ket{\psi_0}$ with a definite configuration of $m$ defects,
that is not necessarily sparse,
has the same entanglement entropy for any region as the ground state,
since it is related to the ground state by $\ket{\psi_0} = P \ket{\Omega_0}$ for some tensor product unitary operator $P$.
Moreover, since the ground state is defined by a commuting set of Pauli operators,
the nonzero Schmidt coefficients with respect to the region $M$ are all the same.
There are precisely $2^{S_0}$ nonzero Schmidt coefficients, where $S_0 = S(M,\ket{\psi_0})/\log 2$.
We consider the Schmidt decomposition of an arbitrary superposition $\ket{\psi_0}$ of the states within the band $P^{(m)}_0$
with respect to the region $M$ and its complement.
Since there are at most $\mathcal O (\mathrm{Vol}~M)^m$ states with different defect configurations in $M$,
the Schmidt rank is at most $\mathcal O ( (\mathrm{Vol}~M)^m 2^{S_0})$.
The topological degeneracy does not enter to this upper bound
since topologically degenerate states have exactly the same reduced density matrix for a ball-like region $M$.
The entanglement entropy is therefore at most $S(M,\ket{\psi_0}) + m \log \mathrm{Vol}~M + O(1)$,
which is an area law.
The result of Ref.~\onlinecite{VanMarienVerstraete2013} states that
for any state $\ket{\psi}$,
the evolved state $U_s \ket{\psi}$ obeys an area law of entanglement entropy whenever $\ket{\psi}$ does.
Since any state $\ket{\psi_s}$ in the band $P^{(m)}_s$
is equal to $U_s \ket{\psi_0}$ for some $\ket{\psi_0}$ in $P^{(m)}_0$,
we have the area law for $\ket{\psi_s}$.

\subsection{Dynamical properties}

Now we are in a position to derive the main result, namely Eq.~\eqref{eq:rough-result}.
The key insight is that the local Hamiltonian has small matrix elements
between the locally gapped state $\ket{\psi_s}$ and the other states.
To formalize this, let us fix a basis of the band subspace $P^{(m)}_s$ consisting of
\[
\tilde{ \ket{ 1}} =\ket{\psi_s} = U_s \ket{\psi_0},  \quad \tilde{ \ket{ q}} = U_s \ket{q} ~(q > 1)
\]
where $\ket{q}$ are energy eigenstate of $H_0$ with definite defect configurations.
Here $\ket{\tilde{1}}$ is approximately locally gapped, but $\ket{\tilde{q}}$ may not be so.
In the Appendix, we show that
\begin{equation}
\sum_{q \neq 1} \abs{ \bra{q} H_s  \ket{\psi_s} } \le g(d)
\label{eq:lemma}
\end{equation}
where $g$ is a rapidly decaying function.
Since the matrix element is so small,
the time evolution can be ``simulated'' by neglecting those small matrix elements.
Indeed, let $Q$ be a hermitian matrix acting within $P_s^{(s)}$,  defined as
\begin{equation}
 \tilde{\bra u} Q \tilde{\ket v} =
 \begin{cases}
  0 & \text{if  $u=1\neq v $ or $u \neq 1 = v$, }\\
  \tilde{\bra u} H_s \tilde{\ket v} & \text{otherwise.}
 \end{cases}
\end{equation}
Then, Eq.~\eqref{eq:lemma} implies that
\begin{equation}
\| P_s(Q - H_s)P_s \| \le g(d) ,
\end{equation}
where we used the fact that the operator norm is at most the sum of absolute values of the matrix elements.
This in turn implies that
\begin{align*}
 &\| e^{-itA} - e^{-it B} \|
 = \left\| 1 - e^{i t A}e^{-i t B} \right\| \\
 &=
 \left\| \int_0^t \rd w \partial_w \left( e^{i w A}e^{-i w B} \right) \right\| \\
 &\le
 \int_0^{|t|} \rd w  \left\|  e^{i w A}( A - B ) e^{-i w B}  \right\| \le |t| g(d)
\end{align*}
where $A=P_s H_s P_s$ and $B= P_s Q P_s$,
from which we conclude that
\[
\abs{ \bra{\psi_s} e^{-itH_s} \ket{\psi_s} } \ge 1 - |t| g(d) .
\]
This is Eq.~\eqref{eq:rough-result}.

The fact that these localized states remain almost invariant under time evolution
is closely related to the existence of the local integrals of motion.
The definition of local integrals of motion in our context is a bit more relaxed
than those that are discussed in the context of many-body localization.
Since we are studying a finite-energy subspace,
the integrals of motion of our interest
may not commute with the Hamiltonian on the entire Hilbert space.
In fact, since our system can be translation-invariant,
we should not expect such commutation relation in general.
However, there are local operators that commute with our Hamiltonian within the localized subspace,
e.g., the subspace spanned by the locally gapped states of the perturbed Hamiltonian.

More concretely, let us define $P_\mathrm{loc}$ be the projector onto the localized subspace.
This refers to the linear span of $U_s \ket{q}$,
where $\ket{q}$ describes a state with defects which are separated by a distance $d$.
Define
\begin{equation}
\mathcal I_j(s) = U_s h_j^{(0)} U_s^\dagger
\end{equation}
 where $h_j^{(0)}$ is the local term of the Hamiltonian $H_0$.
This operator is quasi-local due to the locality-preserving property, i.e., Eq.~\eqref{eq:localityPreserving}.

Now let us see whether this operator commutes with the Hamiltonian within the localized subspace.
\begin{align}
 \left\| P_\mathrm{loc} [ H_s, \mathcal I_j(s) ] P_\mathrm{loc} \right\|
 &\le \sum_{a,b} \abs{ \tilde{\bra{a}} [H_s, \mathcal I_j(s)] \tilde{\ket{b}} }\nonumber \\
 &= \sum_{a,b} \abs{(\ell_a-\ell_b) \tilde{\bra{a}} H_s \tilde{\ket{b}} }\nonumber \\
 &\le \sum_{a \neq b} 2\abs{ \tilde{\bra{a}} H_s \tilde{\ket{b}}} \le 2(\dim P_\mathrm{loc}) g(d) \nonumber \\
 &\le 2L^{3m} g(d). \label{eq:result-IntMotion}
\end{align}
where $L$ is the linear system size.
The first inequality is the triangle inequality,
the second one follows from $\| \mathcal I_j(s) \| = 1$,
and the third one is from Eq.~\eqref{eq:lemma}.
If we choose $d \sim L^{1-\epsilon}$ for a small $\epsilon \in (0,1)$,
then the upper-bound decays rapidly in the system size, and Eq.~\eqref{eq:fractionlocally gapped} reads
\[
 \frac{\dim P_\mathrm{loc}}{\dim P_s^{(m)}} \ge 1 - L^{-\beta} .
\]
for some $\beta>0$.
Thus, $\mathcal I_j(s)$ is almost an integral of motion for almost all states of the band.
Indeed, the expectation value of $\mathcal I_j(s)$ with respect to locally gapped states is slowly varying.
Let $\ket{\psi}$ be {\em any} state in the subspace $P_\mathrm{loc}$.
\begin{align*}
&\bra{\psi} e^{itH_s} \mathcal I_j(s) e^{-itH_s} \ket{\psi} \\
&=
\bra{\psi} \mathcal I_j(s) \ket{\psi} + \int_0^t \rd w \bra \psi e^{iwH_s}i[H_s, \mathcal I_j(s)] e^{-iwH_s} \ket \psi \\
&=
\bra{\psi} \mathcal I_j(s) \ket{\psi} \pm \int_0^t \rd w \mathcal O\left( \| P_\mathrm{loc} [H_s,\mathcal I_j(s)] P_\mathrm{loc} \| \right) \\
&= \bra{\psi} \mathcal I_j(s) \ket{\psi} \pm t \mathcal O\left( L^\alpha g(d) \right)
\end{align*}
The expectation value of $\mathcal I_j(s)$ stays at the initial value up to a time that grows rapidly in the system size.

\section{Discussion}
There is a considerable amount of work in the literature
regarding the fate of the disorder-driven localized system in the presence of interaction;
see e.g. review articles Ref.~\onlinecite{NandkishoreHuse2015,GogolinEisert2015}.
Here we have taken a different route, and identified a mechanism for localization which is driven by a strong interaction.
By exploiting the fact that the ground state of our model is topologically ordered,
we rigorously showed that the localized energy profile of the majority of the low energy states
remains unchanged under arbitrary perturbations.

We remark that, though our analysis was focused on one particular model,
a similar phenomenon is also observed in other models such as Chamon's spin model%
\cite{Chamon2005Quantum,BravyiLeemhuisTerhal2011Topological}
and other ``cubic code'' models\cite{Haah2011Local,Kim2012qupit,Yoshida2013Fractal}.
In Chamon's model, there exists a string operator that is capable of moving topologically nontrivial excitation.
However, there exists an isolated excitation such that no operator can move it without creating extra excitations.
A sparse configuration of these isolated excitations is a locally gapped state,
and our analysis carries over.

Our work strongly suggests that the prevalent dichotomous view on the role of disorder and interaction needs to be modified.
That is, certain strong enough interactions can drive the system into an exotic topologically ordered phase, in which localization occurs at low energy.

An outstanding open problem is whether a locally gapped state can exist at finite energy density,
and remain stable against perturbations.
At such high energy, the gap between the bands would collapse upon perturbation,
and the machinery of quasi-adiabatic continuation in its present form cannot be applied.

\begin{acknowledgments}
We thank T. Senthil for bringing our attention to this problem,
and also Soonwon Choi for insightful discussions.
JH is supported by Pappalardo Fellowship in Physics at MIT.
IK's research at Perimeter Institute is supported by the Government of Canada through Industry
Canada and by the Province of Ontario through the Ministry of Economic Development
and Innovation.
\end{acknowledgments}

\appendix

\section{Generator of quasi-adiabatic continuation}
The quasi-adiabatic continuation operator($U_s$) obeys the following differential equation:
\begin{equation}
\frac{d}{ds}U_s=i K(s)U_s,
\end{equation}
where $U_0=I$. Here we explicitly write down the generator, namely $K(s)$.
While there can be many different choices of the generator, the specific choice we make is due to Bachmann {\it et al}.%
\cite{BachmannMichalakisNachtergaeleSims2012Automorphic}:
\begin{equation}
K(s) = \int^{\infty}_{-\infty} w_{\gamma}(t) \int ^t_0 \rd u e^{iuH(s)}Y e^{-iuH(s)}.
\end{equation}
Here $w_{\gamma}(t)$ is a normalized and rapidly decaying function
whose Fourier transform is supported on $[-\gamma, \gamma]$.
Such a function is known to exist.
An example is given in Ref.~\onlinecite{BachmannMichalakisNachtergaeleSims2012Automorphic},
which decays faster than $\exp(-c\gamma t/ \log ^2(\gamma t))$ for some $c>0$.
If the energy gap between a band and the rest of the spectrum is bounded from below by $\gamma$,
$K(s)$ generates the unitary flow of the band subspace along the path.
The rapid decay of $w_{\gamma}(t)$ implies that $U_s$ preserves locality.
In particular, it is known that $U_s$ maps a local operator to a quasi-local operator,
i.e., an operator that can be well-approximated by a local operator.%
\cite{Hastings2004,BachmannMichalakisNachtergaeleSims2012Automorphic}

\section{Proof of Eq.~\eqref{eq:lemma}}
Consider the off-diagonal element of the Hamiltonian between two orthogonal states within the band subspace.
\[
 \tilde{\bra q} H_s \tilde{\ket 1} = \sum_i \bra{q} U_s^\dagger h_i U_s \ket{1}
\]
The term $U_s^\dagger h_i U_s$ is quasi-local, and thus can be decomposed as $\sum_{r \ge 0} h_{i,r}'$,
where $h_{i,r}'$ is supported on a ball of diameter $r$ centered at $i$ such that
$\| h_{i,r}' \| \le f(r)$ for a rapidly decaying function $f$. We need to consider two cases.

Firstly, if $\ket{q}$ and $\ket{1}$ have the same defect configuration,
i.e., they are topologically degenerate,
then $h_{i,r}$ must have support across the system; $r \sim L$.
The topological degeneracy given a defect configuration is 4
for the cubic code under the aforementioned boundary conditions.
Thus, the contribution from the topological degeneracy is at most $4f(L)$.

Secondly, suppose $\ket{q}$ and $\ket{1}$ have different defect configurations.
For each $i$, let $j(i,q)$ be the furthest location from $i$ at which
the defect is present, \emph{only} in one of the two states.
Each off-diagonal terms, i.e., $\bra{q} h'_{i,r} \ket{1}$, vanish
unless $h'_{i,r}$ creates or annihilates a defect at $j=j(i,q)$.
In order for this to happen, $r$ has to be larger or equal to the distance between $i$ and $j$.
Therefore, if $\mathrm{dist}(i,j)$ is large, the off-diagonal terms decay rapidly.
If $\mathrm{dist}(i,j)$ is small, a better bound can be obtained as follows.
Since $\ket{1}$ is locally gapped with a distance $d$, $\bra{q} h'_{i,r} \ket{1}$ vanishes for $r<d$.
These two cases cover all the possibilities, and we conclude that the nonzero contributions come from $h'_{i,r}$
with $r \ge R(i,q) = \max( \mathrm{dist}(i,j), d )$:
\[
\abs{\bra{q} U_s^\dagger h_{i} U_s \ket{1}} \le \sum_{r \ge R(i,q) } f(r) =: f_1(R(i,q)).
\]
We now sum the right-hand side over all the states $q$ and terms $h_i$, to obtain Eq.~\eqref{eq:lemma}.

Suppose $m$ is the number of defects in $\ket{1}$.
Let us upper-bound the number of tuples $(i,q)$ such that $\mathrm{dist}( i, j(i,q) ) \le x$ for a given $x$.
The defects of $\ket{q}$, as well as the site $i$, has to be contained in a $2x$-neighborhood of the defects of $\ket{1}$;
otherwise the distance between $i$ and $j$ is larger than $x$.
The number of such configurations is at most $((2x)^3)^m \cdot m(2x)^3 \sim x^{3m+3}$.
Then,
\begin{align*}
 &\sum_{i,q} f_1(R(i,q)) \\
 &= \sum_{i,q: \mathrm{dist}(i,j) \le d } f_1(d) + \sum_{i,q:\mathrm{dist}(i,j) = x > d} f_1(x) \\
 &\le O(1) \sum_{x \ge d} x^{3m+3} f_1(x),
\end{align*}
where in the last line we made use of the fact that the topological degeneracy is finite.
Accounting for the topological degeneracy, we get
\begin{equation}
\sum_{q \neq 1} \abs{ \bra q H_s  \ket 1 } \le 4f(L)+ O(1)\sum_{x \ge d} x^{3m+3} f_1(x) .
\end{equation}
Both $f$ and $f_1$ are rapidly decaying.
Eq.~\eqref{eq:lemma} is proved.

\bibliography{localsuper}
\end{document}